\documentclass[12pt,leqno]{article}
\usepackage{graphicx,amsthm,amsmath,amssymb}
\makeatletter	  
\renewcommand\section{\@startsection {section}{1}{\z@}%
                                   {-3.5ex \@plus -1ex \@minus -.2ex}%
                                   {2.3ex \@plus.2ex}%
                                   {\normalfont\large}}
  \makeatother                                 
 \newtheorem{proposition}{Proposition}
 \newtheorem*{conj}{Conjecture}
\newcommand{\vts}{\mkern1mu}
\newcommand{\ts}{\mkern2mu}		
\newcommand{\U}{\operatorname{\mathsf U}}
\newcommand{\ri}{\mathrm{i}\vts}	
\newcommand{\id}{\operatorname{ id}}
\newcommand{\rd}{\operatorname{d}\!}
\newcommand{\End}{\mathop{\mathsf{End}}}
\newcommand{\Cl}{\mathop{\mathsf{Cl}}}
\newcommand{\bwedge}{ \text{\Large\( \wedge  \)}}

\begin{document}
\begin{center}
{\small This paper is due to appear in the\\  Engelbert Schucking Festschrift 
\\
to be published by Springer-Verlag}\\
\vspace{2cm}

{\large\bf ON COMPLEX STRUCTURES IN PHYSICS}\\
\vspace{1.5cm}

{\sc Andrzej Trautman}\\
\bigskip

Instytut Fizyki Teoretycznej UW\\
Ho\.za 69, 00681 Warszawa, Poland\\
E-mail: {\tt amt@fuw.edu.pl}
\end{center}
\vspace{.5cm}

\begin{abstract}
Complex numbers enter fundamental physics
 in  at least two rather distinct ways.
They are needed in quantum theories to make linear differential 
operators into Hermitian observables. Complex structures appear also, 
through Hodge duality, in 
vector and spinor spaces associated with space-time. This paper 
reviews some of  these notions. Charge 
conjugation in multidimensional geometries and
 the appearance of Cauchy-Riemann structures in Lorentz 
manifolds with a congruence of null geodesics without shear are 
presented in considerable detail.
\end{abstract}

\section{Introduction}
In 1960, Ivor Robinson and I studied a class of solutions of 
Einstein's equa\-tions on a Lorentzian manifold, foliated by a shearfree,
non-rotating and diverging congruence of null geodesics. 
We were surprised to find that, in the coordinate system we were using, 
the vanishing of the Ricci tensor implied that 
two components of the metric tensor satisfied Cauchy-Riemann 
equations \cite{RT1}. 
Since that time I have been interested in the question how and why 
complex numbers and structures appear in physical problems.

Complex numbers have been used in physics 
so much and for so long  that one is taking them, most of the 
time, for granted. Their origin is in `pure' mathematics: they 
appeared, in the XVIth century, 
in connection with solving polynomial equations. They were used 
in solutions of physical problems, such as reflection and diffraction of 
waves,   early in the XIX century. Complex roots of polynomial 
equations 
often appear in physical problems in connection with
 linear differential equations with constant coefficients. 
An early, ingenious use of complex analytic functions was made by 
Arnold Sommerfeld 
in his rigorous solution of the  problem of diffraction of waves on a 
half-plane \cite{So}\footnote{I am grateful to J\"urgen Ehlers for 
this reference.}.  
Roger Penrose, in his twistor theory,  put forward convincing 
arguments in favor of the relevance of holomorphic structures for 
fundamental physics \cite{Pe,PR}. His methods and ideas have been successfully 
used in a variety of mathematical and physical problems. Complex numbers 
and analytic functions permeate 
now all of quantum physics.

Complex numbers are introduced in physical theories in several ways; 
it is not obvious that `all the square roots of \(-1\) are the same'. 
There is the \(\sqrt{-1}\) of quantum mechanics that is `universal' in 
the sense that it appears irrespective of the details of the model 
under consideration. Chiral (Weyl, reduced) spinors in Minkowski space-time are 
also complex, but this property reflects  the signature of its metric 
tensor. Dirac spinors can be restricted to be real, provided one uses 
a metric of signature \((3,1)\); but to write the wave equation of an 
electron, in any signature,
 one has to introduce complex numbers because electromagnetism 
is a gauge theory with \(\U_{1}\) as the structure group. The 
`electromagnetic'  \(\sqrt{-1}\) seems to have a quantum-theoretical 
origin: in the classical theory of a particle of charge \( e \) the 
potential \( A_{\mu} \) appears in expressions such as \( 
p_{\mu}+eA_{\mu} \); in quantum theory this becomes
\[
\ri\partial_{\mu}+eA_{\mu}=\ri (\partial_{\mu}-\ri eA_{\mu}).
\]
and the \( \ri \) next to \( eA_{\mu} \) reflects the nature of the Lie algebra of
\(  \U_{1}  \).

In this article, I  present some thoughts on the origin of the
appearance of complex numbers in 
physics, emphasizing  the geometric, 
rather than the analytic, aspects of the problem. 
After recalling the notion of complex structures in  real 
vector spaces (Section \ref{amt:Cs}), I  show, on a simple example,
 how such structures  may be considered to appear in quantum 
mechanics (Section \ref{amt:QM}). For some signatures of the metric tensor, 
Clifford-Hodge-K\"ahler duality introduces complex structures in spaces of 
spinors and multivectors (Section \ref{amt:CHK}). Charge conjugation is also closely related 
to the appearance of complex numbers in quantum theory; its 
generalization to higher dimensions is described in Section 
\ref{amt:chconj}. In a  final section, influenced by my 
collaboration with W. Kopczy\'nski, P.
 Nurowski and J. Tafel, I describe the geometry underlying 
 shearfree congruences of null geode\-sics, its relation to 
 Cauchy-Riemann structures in three dimensions, and the close 
analogy between optical geometries in Lorentzian 
manifolds and Hermitian geometries
in proper Riemannian manifolds.
\medskip

In 1961, Engelbert Schucking and I spent some time together at Syracuse 
University in Peter Bergmann's group that included also  Dick 
Arno\-witt, Asim Barut,
Art Komar, Ted Newman, Roger Penrose,  Ivor Robinson, Ralph 
Schiller and Mel Schwartz. Since that time, 
I have had the pleasure to  see Engelbert on various 
occasions and to talk with  him 
 on many issues of science and life. These discussions included also 
 the topics touched upon in this text. On one of my visits to New York, Engelbert 
 presented me with a copy of  \cite{Za}, an excellent account of the 
 history of number systems. Several times, 
my family and I enjoyed his very kind hospitality at Washington Square. 
This article is dedicated to Engelbert as a token of my friendship and respect.

\section{Definitions and notation}\label{amt:Cs}

Recall that a {\em complex structure\/} in a real vector space \( W\) is
 a linear automorphism \( J \) of \( W \) such that \( J^{2}=-\id_{W} 
 \); if \( W \) is finite-dimensional, then its dimension is even. A 
 real vector space \( W \) with a complex structure \( J \) can be 
made into a complex vector space in two ways, by defining, for every \( 
w\in W \), either \( \ri w =J(w)\) or \( \ri w=-J(w)\). The 
automorphism \( J \) extends, in an obvious way, to an automorphism
\( J_{\mathbb C} \) of the 
{\em complexification\/} of \( W \), i.e. to the complex vector space
\( \mathbb C\otimes W \). This space can be decomposed into the direct sum,
\begin{equation}\label{amt:1}
\mathbb C\otimes W=W_{+}\oplus W_{-},\;\;\text{where}\;\;
W_{\pm}=\{w\in \mathbb C\otimes W:  J_{\mathbb C}(w)=\pm\ri w\}.
 \end{equation}
Considered as a complex vector space, \( W \) is isomorphic to  \( 
W_{+} \) or \( W_{-} \), depending on whether the multiplication by \( \ri 
\)  in \( W \) is defined as  \( \ri w=J(w) \) or \( \ri w =-J(w)
\), respectively. Note that if \( W \) is of real dimension \( 2n \), then 
its complex dimension is \( n \). Every complex vector space of 
dimension \( n \)  can be
`realified', i.e. considered as a real vector space of real dimension 
\(2n\). Such a realification has a natural complex structure.

Assume now that the real vector space \( W \) has a (generalized) scalar product, 
i.e. a map \( g:W\times W\to \mathbb R \) which is bilinear, 
symmetric and non-degenerate. The scalar product \( g \) extends to a \( 
\mathbb C \)-bilinear scalar product \( g_{\mathbb C} \) on \( 
\mathbb C\otimes W \). If the complex structure \( J \) in \( W 
\) is orthogonal with respect to \( g \), i.e. if \( 
g(J(w_{1}),J(w_{2}))=g(w_{1},w_{2}) \) for every \( w_{1},w_{2}\in W 
\), then the vector spaces \( W_{+} \) and \( W_{-} \) are both 
totally null (isotropic) with respect to \( g_{\mathbb C} \). 
The vector space \( W \), considered as a 
complex vector space such that \( \ri w=J(w) \) has a Hermitian scalar 
product \( h:W\times W\to \mathbb C \) defined by
\begin{equation}\label{amt:defh}
h(w_{1},w_{2})=g(w_{1},w_{2})+\ri g(J(w_{1}),w_{2}),\quad 
w_{1},w_{2}\in W,
\end{equation}
so that \( h(w_{1},\ri w_{2}) =\ri h(w_{1},w_{2})\), \( 
\overline{h(w_{1},w_{2})}=h(w_{2},w_{1}) \) and \( h(w,w)=g(w,w) \).
\medskip

Consider  now a {\em complex, finite-dimensional vector space\/}  \( S \). 
Its (complex)
dual  \( S^{*} \) consists of all \( \mathbb C \)-linear maps \( 
s':S\to\mathbb C \); it is often convenient to denote here the value of
\( s' \) on \( s\in S \) by \( \langle s',s\rangle  \). If \( 
f:S_{1}\to S_{2} \) is a \( \mathbb C \)-linear map of complex vector 
spaces, then the dual (transposed) map \( f^{*}:S_{2}^{*}\to S_{1}^{*}\) 
is defined by \( \langle f^{*}(s'), s\rangle =\langle s',f(s)\rangle 
\) for every \( s\in S_{1} \) and \( s'\in  S_{2}^{*}\). The spaces \( 
S^{**} \) and \( S \) can be identified.  A map \( 
h:S_{1}\to S_{2} \) is said to be {\em antilinear\/} (semi-linear) if it is 
\( \mathbb R \)-linear and \( h(\ri s)=-\ri h(s) \) for every \( s\in 
S_{1} \). The {\em complex conjugate\/} \( \bar{S} \) of a complex 
vector space \( S \) is the complex vector space of all antilinear 
maps of \( S^{*} \) into \( \mathbb C \); there is a canonical 
antilinear isomorphism \( S\to \bar{S} \), \( s\mapsto\bar{s} \), 
given by \( \langle \bar{s},s' \rangle=\overline{\langle 
s',s\rangle}\). With every linear map \( f:S_{1}\to S_{2} \) there is 
associated the linear map \( \bar{f}:\overline{S_{1}}\to \overline{S_{2}} \)
defined by \( \bar{f}(\bar{s})=\overline{f(s)} \) for \( s\in S_{1} 
\); the map \( f\mapsto \bar{f} \) is antilinear. If \( g:S_{2}\to S_{3} 
\) is another linear map, then  \((g\circ f)^{*}=f^{*}\circ g^{*} \) 
and \( \overline{g\circ f}=\bar{g}\circ\bar{f} \).  One often writes \( 
gf \) instead of \( g\circ f \).

All manifolds and maps are assumed to be smooth. Einstein's summation 
convention over repeated indices is used.  If \( L \) is a vector
bundle over a manifold \( M \), then \( \varGamma(L) \) denotes the module 
of sections of \( L\to M \). The zero bundle is denoted by \( 0 \). 
The tangent and cotangent bundles of \( M \) are denoted by \( TM \) 
and \( T^{*}M \), respectively.
The contraction of a vector (field) \( v \) 
with a \( p \)-form \( \omega \) is the \( (p-1) \)-form \( 
v\lrcorner\vts\omega \) given by its value on the vectors \( 
v_{2},\dots,v_{p} \), \( \;(v\lrcorner\vts\omega)(v_{2},\dots,v_{p})=
\omega(v,v_{2},\dots,v_{p}) \). If \( g \) is a scalar product on a 
vector space \( V \) and \( v\in V \), then \( g(v)\in V^{*} \) is 
defined by \( v'\lrcorner\vts g(v)=g(v',v) \) for every \( v'\in V \).
The exterior  differential of a form \( \omega \) is denoted by \( 
\rd \omega \).

 \section{A complex structure defined by differentiation}\label{amt:QM}
 
 The usual argument for complex numbers in quantum mechanics, in a 
 simplified form,  runs as 
 follows:  differential operators such as \( 
 \partial/\partial x \),  because of their relation to translations, 
 are needed to represent components of momentum; to make them (formally)
 self-adjoint, one has to multiply  by \( \ri \). One can 
 reformulate this argument into a statement about the appearance of 
 a complex structure in the vector space of wave functions, initially 
 considered as a real vector (Hilbert) space. The key observation is 
 that the Laplacian on a compact, proper Riemannian manifold is 
 a {\em negative\/} operator.
 
 To illustrate this argument on a simple example, and make it
  explicit, consider the infinite-dimensional 
 real Hilbert space  \( L^{2}_{\mathbb R}(\mathbb S_{1}) \) of square-integrable
 functions on the circle \( \mathbb S_{1} \). Let 
   \( x \) be a coordinate on the circle, \( 0\leqslant x\leqslant 2\pi \).  
The scalar product of two  functions
  \( \varphi,\psi:\mathbb S_{1}\to\mathbb R \), is given by
 \[
 g(\varphi,\psi)=\int_{0}^{2\pi}\varphi(x)\psi(x)\rd x
 \]
 so that  \( g(\varphi,\varphi)\geqslant 0\). 
 Let \( W \) be the vector subspace of \( L^{2}_{\mathbb R}(\mathbb S_{1}) \) 
 containing all functions orthogonal to the constants on the circle,
 \[
 W=\{\varphi\in\ L^{2}_{\mathbb R}(\mathbb S_{1} ): 
 \int_{0}^{2\pi}\varphi(x)\rd x=0\}.
 \]
 Smooth functions in \( W \) constitute a dense subspace of that 
 space; for every 
 two such functions \( \varphi \) and \( \psi \) one has
 \[
 g(\varphi', \psi)=-g(\varphi,\psi'),
 \]
 where \( \varphi'(x)= \rd \varphi(x)/\rd x\).
 The operator \( \rd^{2}/\rd x^{2} \) is (formally) 
 self-adjoint and negative on \( W \): if \( \varphi \) is smooth and 
 \( \varphi\neq 0 \), then
 \[
 g(\varphi, \varphi'' )=-g(\varphi',\varphi')<0.
 \]
 The set of eigenfunctions of \( \rd^{2}/\rd x^{2} \),
 \[
 \{\cos kx,\;\sin kx\},\quad k=1,2,\dots,
 \]
  is a basis in \( W \). The operator \( -\rd^{2}/\rd x^{2} \)
  has only positive eigenvalues; as such it has  a 
  unique positive square root \( X \),
  i.e. a (formally) self-adjoint operator with positive eigenvalues such 
  that \( X^{2}=-  \rd^{2}/\rd x^{2} \). The operator \( X \) 
  in \( W \), 
  which may be characterized by its action on the basis vectors,
  \[
  X(\cos kx)=k\cos kx,\quad X(\sin kx)=k\sin kx,
  \]
  is {\em invertible\/} and {\em commutes\/} with the operator
   \( \rd/\rd x\). Therefore, the linear operator
   \[
   J=X^{-1}\circ\frac{\rd} {\rd x}\quad\text{satisfies}\quad
   J^{2}=-\id_{W}
   \]
  and defines a complex structure on \( W \). Introducing the complex 
  vector spaces \( W_{\pm} \), as in the previous section, one obtains
  \[
  X=\mp \ri \frac{\rd} {\rd x}\quad\text{on}\quad W_{\pm}.
  \]
  
  \section{Complex structures associated with\\ pseudo-Euclidean vector 
  spaces}\label{amt:CHK}
  
  Let \( V \) be a real, \( m \)-dimensional vector space with a 
  scalar product \( g \) of signature \( (k,l) \), \( k+l=m \). The 
  Clifford algebra associated with the pair \( (V,g) \) is denoted by 
  \( \Cl_{k,l} \).   The  algebra 
  is generated by \( V \); by declaring the elements of \( V  \) to 
  be {\em odd}, one defines a \( \mathbb Z_{2} \)-grading of  \( 
  \Cl_{k,l} \):  one writes \( \Cl_{k,l}^{0}\to \Cl_{k,l} \) to 
  emphasize this grading and exhibit the  even subalgebra \( 
  \Cl_{k,l}^{0}\). The {\em degree\/} \( \deg a \) of an even (resp., 
  odd) element \( a\in\Cl_{k,l} \) is 0 (resp., 1). 
   Recall that if \( \mathcal A \) and \( \mathcal B 
  \) are \( \mathbb Z_{2} \)-graded algebras, then multiplication in 
  their {\em graded product\/} \( \mathcal A\otimes_{\rm gr}\mathcal B 
  \) is defined, for homogeneous elements \( a'\in \mathcal A \) and
  \( b\in\mathcal B \), by \( (a\otimes b)(a'\otimes 
  b')=(-1)^{\deg b \deg a'}aa'\otimes bb' \).
 For every \( k,l\in\mathbb N \) the algebras \( \Cl_{k,l} \) 
  and \( \Cl^{0}_{k,l+1} \) are isomorphic. 
  Denote by \( \mathbb R(N) \) the algebra of real \( N \) by \( N 
  \) matrices.  For every  algebra \( \mathcal A \) over
   \( \mathbb R \), put \( 2\mathcal A 
  =\mathcal A\oplus \mathcal A\) and \( \mathcal A(N)= \mathcal 
  A\otimes \mathbb R(N)\). Every Clifford algebra \( \Cl_{k,l} \) is 
  isomorphic to one of the following algebras:
 \(  \mathbb R(2^{p}) \), \( \; \mathbb C(2^{p}) \), \( \; \mathbb H(2^{p}) \),
 \( \;2\mathbb R(2^{p}) \), \( \; 2\mathbb H(2^{p}) \), \( \; p\in\mathbb N \).
 Recall the {\em Chevalley theorem\/}: \( \Cl_{k,l}\otimes_{\rm 
 gr}\Cl_{k',l'}=\Cl_{k+k',l+l'} \) and the isomorphisms:
 \( \Cl_{k+4,l}=\Cl_{k,l+4} \), \( 
 \Cl_{k+1,l+1}=\Cl_{k,l}\otimes\mathbb R(2) \).
  Two Clifford algebras, \( \Cl_{k,l} 
 \) and \( \Cl_{k',l'} \),  are said to be of the same {\em type\/} if
 \( k+l'\equiv k'+l\bmod 8 \).
 When grading is taken into account, there are  eight  types
 of  Clifford algebras; with respect to graded tensor multiplication 
 the set of these eight types forms a group (the {\em Brauer-Wall 
 group\/} of \( \mathbb R \)) isomorphic to \( \mathbb Z_{8} \); for 
 this reason, the algebras  are 
 conveniently represented on the {\em spinorial clock\/} \cite{BT}:

  \begin{center}
\includegraphics
[scale=.8,bb= 52 548 262 758,clip]
{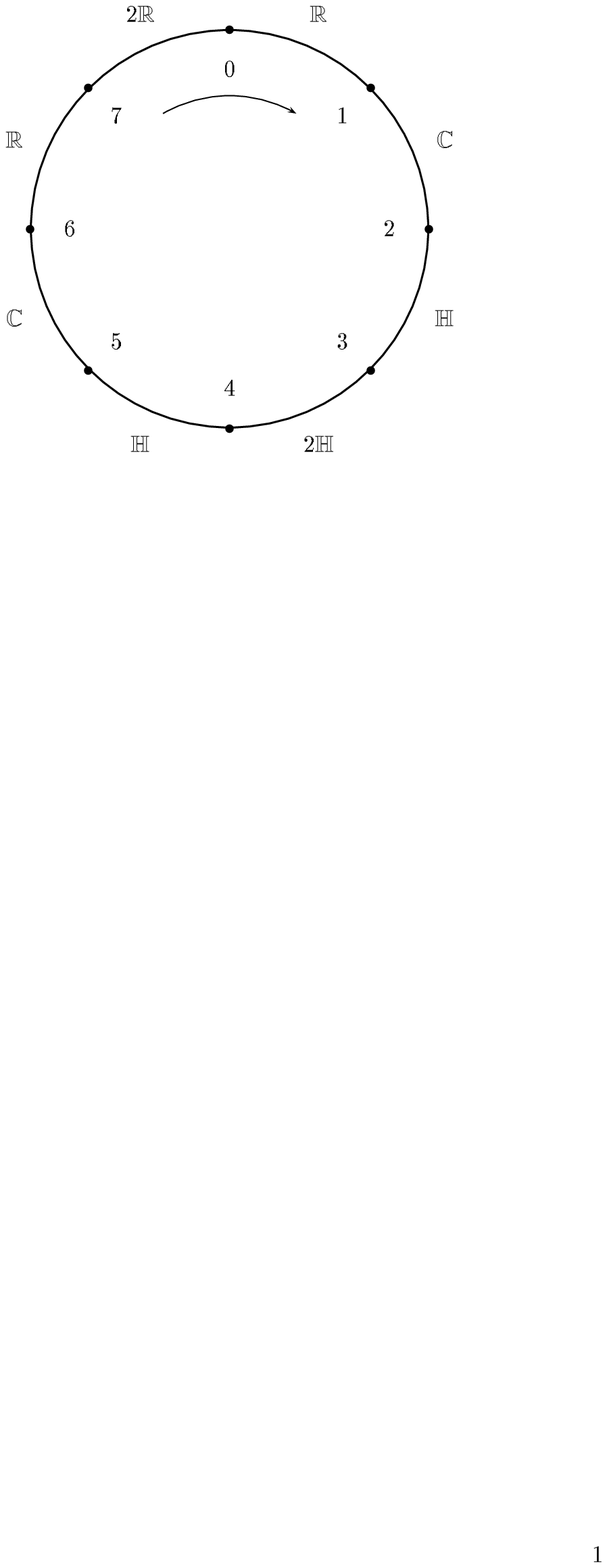}
\vspace{2ex}

\begin{minipage}{20pc}
{\small
The  spinorial clock can be used to find the structure of real 
Clifford algebras. To determine \( \Cl_{k,l}^{0}\to\Cl_{k,l} \), 
compute the corresponding {\em hour\/} \( h\in\{0,\dots,7\} \), \( 
l-k=h+8r \), \( r\in\mathbb Z \). Read 
off the sequence \( \mathcal A_{h}^{0} 
\stackrel{h}{\longrightarrow}\mathcal A_{h}\)
 from the clock. If \( \dim_{\mathbb R}\mathcal A_{h}=2^{\nu_{h}} \),
 then \( \Cl_{k,l}=\mathcal A_{h}(2^{\tfrac{1}{2}(k+l-\nu_{h})}) \), 
 etc.
 The algebra \( \mathcal A_{h}\otimes_{\rm gr}\mathcal A_{h'} \) is of 
 the same type as \( \mathcal A_{h+h' \bmod 8} \).}
\end{minipage}
\end{center}
\medskip

As a vector space, the algebra
  \( \Cl_{k,l} \) is isomorphic, in a natural way, to the vector space
 \( \bwedge V \) underlying the exterior algebra of \( V \). This 
  isomorphism
  \[
  \kappa:{\Cl}_{k,l}\to 
 \bwedge V
  \]
  is characterized by \( \kappa(1)=1\)  and 
  \[ 
  \kappa(va)= 
  v\wedge\kappa(a)+g(v)\lrcorner\vts \kappa(a)\quad\text{for every \( v\in 
  V\subset{\Cl}_{k,l} \) and \( a\in{\Cl}_{k,l} \)}.
  \]
 It respects the \( \mathbb Z_{2} \)-grading of the vector spaces in 
  question.
 
An orthonormal frame \( (e_{\mu}) \) in \( V \subset \Cl_{k,l}\) satisfies
  \( e_{\mu}e_{\nu}+e_{\nu}e_{\mu}=0 \) for \( \mu\neq\nu \),
  \( e_{\mu}^{2}=1 \) for \( \mu=1,\dots, k \), and \( =-1 \) for
  \( \mu=k+1,\dots,m \). The square of the {\em volume element\/}
  \(\eta=e_{1}\dots e_{m}\)
  is  \(
  \eta^{2}=(-1)^{\tfrac{1}{2}(l-k)(l-k+1)}\). Hodge duality, as 
  defined by K\"ahler, is given by
  \[
  \star \kappa(a)=\kappa(\eta a),\quad\text{where}\quad a\in{\Cl}_{k,l}.
  \]
  
 Whenever \( \eta^{2}=-1 \), there is a natural complex structure
  in the real vector space \( \bwedge V \). Therefore,
 
 \noindent\  (i) if \( V=\mathbb R^{2n} \) and \( l-k\equiv 2\bmod 4\), then \( \star 
 \) defines a complex structure in \( \bwedge^{n}\mathbb R^{2n} \);
 
 \noindent (ii) if \( V=\mathbb R^{2n+1}\) and \( l-k\equiv 1\bmod 4 
 \), then \( \star \) defines a complex structure in \( 
 \bwedge^{n}\mathbb R^{2n+1}\oplus\bwedge^{n+1}\mathbb R^{2n+1} \).
 \smallskip
  
  If \( m=k+l \) is even, \( m=2n \), then the algebra \( \Cl_{k,l} \) 
  is central simple and has one, up to equivalence, representation 
  \begin{equation}\label{amt:g}
  \gamma:{\Cl}_{k,l}\to\End S 
  \end{equation}
   in a complex, \( 2^{n} 
  \)-dimensional space of Dirac spinors. In particular, the contragredient 
  representation  and the complex conjugate 
   representation  are each equivalent 
  to \( \gamma \). In terms of the Dirac matrices (endomorphisms of \( 
  S \)), \( 
  \gamma_{\mu}=\gamma(e_{\mu}) \), this equivalence may be
   expressed by the equations
   \begin{equation}
   	\gamma_{\mu}^{*}=B\gamma_{\mu}B^{-1},\quad
   	B:S\to S^{*},
   	\label{amt:defB}
   \end{equation} 
   and
   \begin{equation}
   	\overline{\gamma_{\mu}}=C\gamma_{\mu}C^{-1},\quad
   	C:S\to\bar{S}.
   	\label{amt:defC}
   \end{equation}
The intertwining isomorphisms \( B \) and \( C \) 
are defined up to multiplication by non-zero complex numbers. 
 The matrix 
 \[
  \gamma_{2n+1}=\gamma_{1}\dots \gamma_{2n}
  \]
 anticommutes with \( \gamma_{\mu} \) for \( \mu=1,\dots,2n \).
 The representation \( \gamma \), restricted to \( \Cl^{0}_{k,l} \), 
 decomposes into the direct sum \( \gamma_{+}\oplus\gamma_{-} \) of 
 representations,
 \[
 \gamma_{\pm}:{\Cl}_{k,l}^{0}\to \End S_{\pm},
 \] 
 in the spaces \( S_{+} \) and \( S_{-} \) of Weyl spinors of 
 opposite  {\em chirality\/},
 \[
 S_{\pm}=\{\varphi\in S:\gamma_{2n+1}\varphi=\pm\iota\varphi\},
 \]
 where 
 \begin{equation}\label{amt:iota}
\iota=1 \;\;\text{for}\;\;  \eta^{2}=1 \;\;\text{and}\;\;  \iota=\ri 
\;\;\text{for}\;\;  \eta^{2}=-1. 
\end{equation}
 The relevant properties of  
\( B \), \( C \) and \(  \gamma_{2n+1}\) can be summarized in (see, e.g., \cite{BT})
\begin{proposition}\label{amt:prop1}
If \( k+l=2n \), then
\begin{align}
 B^{*}=(-1)^{\tfrac{1}{2}n(n-1)}B,\\
 	\gamma_{2n+1}^{*}=(-1)^{n}B\gamma_{2n+1}B^{-1}\\
 	\overline{\gamma_{2n+1}}=C\gamma_{2n+1}C^{-1}.
 	\label{amt:g5}
 \end{align}
One can normalize the intertwining isomorphisms defined by 
{\rm (\ref{amt:defB})} and {\rm (\ref{amt:defC})} so that
\begin{equation}
 \bar{C}C=(-1)^{\tfrac{1}{8}(l-k)(l-k+2)} \label{amt:CC}
 \end{equation}
 and
 \begin{equation}
 \bar{B}C=\bar{C}^{*}B^{*}.  \label{amt:BC}
 \end{equation}
 \end{proposition} 
 The proof of (\ref{amt:BC})  is based on the observation that 
 \(\bar{B}C(\bar{C}^{*}B^{*})^{-1}  \) is in the commutant of the 
 irreducible representation \( \gamma \).
 
  From the spinorial clock one obtains, for every \( p\in\mathbb N\),
   the isomorphism
  \[
  {\Cl}_{3+p,1+p}=\mathbb R(2^{2+p}).
  \]
  Since in this case \( \eta^{2}=-1 \), there is a complex structure \( 
  J=
  \gamma(\eta) \) in the real space of Dirac-Majorana spinors
  \( W=\mathbb R^{2^{2+p}} \). Defining, as in (\ref{amt:1}), the 
  complex, \( 2^{1+p} \)-dimensional spaces \( W_{+} \) and \( W_{-} 
  \), one sees that they can be identified with the two spaces of Weyl
  spinors of opposite chirality.
  
 \section{Charge conjugation}\label{amt:chconj}
 
 Charge conjugation is intrinsically connected with the equivalence 
 of the representations \( \gamma \) and \( \bar{\gamma} \). 
 The notion of charge conjugation, 
 defined originally by physicists for spinors associated with Minkowski space, 
 admits a generalization to higher dimensions \cite{BT}. In 
 view of some controversy surrounding this generalization \cite{Lo}, I present 
 it here, in considerable detail, for the case of an even-dimensional, 
 flat space-time with a metric of 
 signature \( (2n-1,1) \). 
 
 Consider first the general case of \( k+l \) even, \( k+l=2n \); given a representation
(\ref{amt:g})
 of the algebra \( \Cl_{k,l} \) in a complex vector space \( S \) of Dirac spinors, one 
 defines {\em charge conjugation} to be
 the antilinear map \( S\to S \), 
 \begin{equation}
 	\varphi\mapsto \varphi_{c}=C^{-1}\bar{\varphi}.
 	\label{amt:defpsic}
 \end{equation}
 If \( \varphi \) is a Weyl spinor, \( 
 \gamma_{2n+1}\varphi=\pm\iota\varphi \), then \( \varphi_{c} \) is 
 also such a spinor and its chirality is the same as (resp., opposite to) 
 that of \( \varphi \) if \( \eta^{2}=1 \) (resp., if \( \eta^{2}=-1 \)).
  If \( \bar{C}C=\id_{S} \), then the map
   \( \varphi\mapsto\varphi_{c} \) is involutive, 
  \( (\varphi_{c})_{c}=\varphi \), and there is the real vector space
  \[
  S_{\mathbb R}=\{\varphi\in S:\varphi_{c}=\varphi\}
  \]
   of Dirac-Majorana spinors. Charge conjugation is {\em not\/}, 
   however, restricted to that case. If \( \gamma_{\mu} \) are the Dirac matrices 
  corresponding to a representation of \( \Cl_{k,l} \), then the 
  matrices \( \ri\gamma_{\mu} \) correspond to a representation of 
  \( \Cl_{l,k} \). 

 Assume now that the signature is Lorentzian, \( k=2n-1 \) and
 \( l=1 \). In view of the previous remark, the case of signature \( 
 (1,2n-1) \) can be easily reduced to the one under 
 consideration. The properties of the intertwiner \( C \), described 
 in Prop. \ref{amt:prop1}, are now expressed by the equation
 \begin{equation}
 	\bar{C}C=(-1)^{\tfrac{1}{2}(n-1)(n-2)}\id_{S}.
 	\label{amt:Cn}
 \end{equation}
 The Dirac equation for a 
 particle of mass \( m \) and electric charge \( e \) can be written as
 \begin{equation}
 	\gamma^{\mu}(\partial_{\mu}-\ri eA_{\mu})\psi=m\psi,
 	\label{amt:Dirac}
 \end{equation}
 where \( \psi:\mathbb R^{2n}\to S \) is the wave function of the 
 particle and 
 \( A_{\mu} \), \( \mu=1,\dots, 2n, \) are the (real) components of the vector 
 potential of the electromagnetic field. 
 For a free particle (\( A_{\mu}=0 \))  one can  consider a solution 
 of (\ref{amt:Dirac}) equal to a constant spinor times \( \exp \ri p_{\mu}x^{\mu} \); 
 the Dirac equa\-tion then implies  that the momentum vector 
 \((p_{\mu})\) is time-like: \( p_{2n}^{2}=p_{1}^{2}+\dots 
 +p_{2n-1}^{2}+m^{2} \).
 The {\em charge conjugate 
 wave function\/} \( \psi_{c}: \mathbb R^{2n}\to S \) is defined by
 \( \psi_{c}(x)=\psi(x)_{c} \) for every \( x\in\mathbb R^{2n} \).
 
 \begin{proposition}
If  \(  \psi:\mathbb R^{2n}\to S  \) is  a wave function, then

\noindent\  {\rm (i)}  the vector field of current defined by 
\begin{equation}\label{amt:j}
 j^{\mu}(\psi)=\ri^{n+1}\langle 
 B\gamma_{2n+1}\psi_{c},\gamma^{\mu}\psi\rangle,\;\;\mu=1,\dots,2n,
 \end{equation}
 is {\em real} and invariant with respect to the replacement of \( \psi \) 
 by \( \psi_{c} \),
 \begin{equation}
 j^{\mu}(\psi_{c})= j^{\mu}(\psi); \label{amt:jpsi}
 \end{equation}
 
 \noindent {\rm (ii)} if \( \psi \) is a solution of the Dirac equation 
 {\rm (\ref{amt:Dirac})}, then the current is conserved,  
 \begin{equation}
 \partial_{\mu}j^{\mu}(\psi)=0, \label{amt:ccons}
  \end{equation} 
 and the charge conjugate wave 
 function satisfies the Dirac equation for a particle of charge \( -e 
 \),
  \begin{equation}
 	\gamma^{\mu}(\partial_{\mu}+\ri eA_{\mu})\psi_{c}=m\psi_{c}.
 	\label{amt:cDirac}	
 \end{equation}
 \end{proposition}
 The proof  of part (i) the Proposition consists of simple, algebraic 
 transformations, making use of equations  (\ref{amt:defB}), 
 (\ref{amt:defC}), (\ref{amt:defpsic}),  (\ref{amt:Cn})  and Prop. \ref{amt:prop1}.   
 Complex conjugating both 
 sides  of 
  (\ref{amt:Dirac}), multiplying 
  the resulting equa\-tion  by \( C^{-1} \) on the left and 
  using (\ref{amt:defC}) and (\ref{amt:defpsic}),
  one obtains that \( \psi_{c} \) satisfies (\ref{amt:cDirac}); it is 
  then easy to check that (\ref{amt:ccons}) holds.

 These simple observations are  valid irrespective 
 of whether the algebra \( \Cl_{2n-1,1} \) 
 is real (\( \bar{C}C=\id_{S} \);\, \( n\equiv 1\;\text{or}\;2\bmod 4 \)) or
 quaternionic  (\( \bar{C}C=-\id_{S} \);\, \( n\equiv 
 0\;\text{or}\;3\bmod 4 \)). 
 Charge conjugation is not related to the existence of Majorana 
 spinors: even if the algebra \( \Cl_{2n-1,1} \) is {\em real}, one has to 
 use {\em complex\/} spinors to write the Dirac equation for a 
 charged particle interacting with an electromagnetic field. The 
 invariance of the current under the replacement of \( \psi \) by \( 
 \psi_{c} \), expressed by (\ref{amt:jpsi}), reflects the classical 
 (or rather: first-quantized) 
 nature of the Dirac equation under consideration here. Upon second 
 quantization, the wave function is replaced by an {\em 
 anticommuting\/}, spinor-valued field \( \varPsi \);  anticommutativity 
 of \( \varPsi \) and \( \varPsi_{c} \)  provides 
 a change of sign, so that (\ref{amt:jpsi}) is replaced by 
 \( j^{\mu}(\varPsi_{c})=- j^{\mu}(\varPsi)\).\footnote{I thank 
 Engelbert Schucking for a discussion on this aspect of charge 
 conjugation. See also Appendix I in \cite{Th}.}
 \smallskip
 
 As an  example, consider the case of dimension 8:
  one has \( \Cl_{7,1}=\mathbb H(8) \)
 and the space of Dirac spinors is complex 16-dimensional. Let
 \[
 \sigma_{x}=\begin{pmatrix}0&1\\1&0\end{pmatrix},\quad
 \sigma_{y}=\begin{pmatrix}0&-\ri\\\ri &0\end{pmatrix},\quad
 \sigma_{z}=\begin{pmatrix}1&0\\0&-1\end{pmatrix},\quad
 I=\begin{pmatrix}1&0\\0&1\end{pmatrix}
 \]
 be the  Pauli matrices.
  One can take, in this case, a representation such that 
 \begin{align*}
 \gamma_{1}&=\sigma_{x}\otimes I\otimes I\otimes I,\quad &
 \gamma_{2}&=\sigma_{y}\otimes\sigma_{y}\otimes I\otimes I,\\
 \gamma_{3}&=\sigma_{y}\otimes\sigma_{x}\otimes\sigma_{y}\otimes 
 I,\quad &
\gamma_{4}&=
 \sigma_{y}\otimes\sigma_{x}\otimes\sigma_{x}\otimes\sigma_{y},\\
 \gamma_{5}&=\sigma_{y}\otimes\sigma_{x}\otimes\sigma_{z}\otimes\sigma_{y},
 \quad &
 \gamma_{6}&=\sigma_{y}\otimes\sigma_{z}\otimes 
 I\otimes\sigma_{y},\\
 \gamma_{7}&=\sigma_{y}\otimes\sigma_{z}\otimes\sigma_{y}\otimes\sigma_{x},
 \quad &
 \gamma_{8}&=\ri \sigma_{y}\otimes\sigma_{z}\otimes\sigma_{y}\otimes
 \sigma_{z}, 
 \end{align*}
 and
 \[
 C=\sigma_{x}\otimes\sigma_{z}\otimes\sigma_{y}\otimes\sigma_{z}.
 \]
 Note that the algebra \( \Cl_{7,1}^{0}\to \Cl_{7,1}\) is of the same 
 type as the algebra \( \Cl_{1,3}^{0}\to \Cl_{1,3}\).
   
  \section{CR structures associated with integrable optical geometries}
In this section, intended to `explain' the appearance of 
Cauchy-Riemann structures in the process of solving Einstein's 
equations for special Lorentz metrics, I restrict myself to 
four-dimensional Riemannian manifolds. 

Consider first a Lorentz manifold, i.e. a Riemannian manifold \( M \) with a 
metric tensor field \( g \) of signature \( (3,1) \). Assume that \( M \) is space and time 
oriented and  that there is given on \( M \) a bundle \( K\subset TM \)
 of null lines; the flow  generated  by \( k\in \varGamma(K)\)  has 
null curves (rays) as trajectories. Since the fibres of \( K \) are null, the 
bundle
\[
K^{\perp}=\{u\in TM:g(k,u)=0\;\;\text{for every}\;\;k\in\varGamma(K)\}
\]
contains \( K \) and there is the exact sequence of homomorphisms of 
vector bundles,
\[
0\to K\to K^{\perp}\to K^{\perp}/K\to 0.
\]
The fibres of \( K^{\perp}/K \) are 2-dimensional and have a 
positive-definite scalar product induced by \( g \): they are the {\em 
screen spaces\/} of the `optical' geometry of rays \cite{EK,Nu,PR,RT}. Space and 
time orientation of \( M \), together with the conformal structure of 
the screen spaces, induce a complex structure 
\( J \) in the fibres of \( K^{\perp}/K  \).  There is a natural 
extension \( J_{\mathbb C} \) of \( J \) to the complexified bundle
\( \mathbb C\otimes (K^{\perp}/K) \); the latter bundle can be 
identified with \( (\mathbb C\otimes K^{\perp})/(\mathbb C\otimes K) \).
For every \( n\in  \mathbb C\otimes K^{\perp}\),  let  \( n+K\in
(\mathbb C\otimes K^{\perp})/(\mathbb C\otimes K) \) denote
 the coset space  containing \( n \).  The  vector bundle
\begin{equation}
	N=\{n\in\mathbb C\otimes K^{\perp}:J_{\mathbb C}(n+K)=\ri n+K\}
	\label{amt:N}
\end{equation}
is a subbundle of \( \mathbb C\otimes TM \); its fibres are complex, 
totally null planes, \( N^{\perp}=N \), and
\begin{equation}\label{amt:NK}
N\cap \bar{N}=\mathbb C\otimes K,\quad N+\bar{N}=\mathbb C\otimes 
K^{\perp}.
\end{equation}
A totally null, complex plane  bundle \( N \) can be also considered 
in other possible  signatures 
(namely, \( (4,0) \) and \( (2,2) \)) of \( g \) on a  4-manifold.
\smallskip

If \( g \) is a {\em proper Riemannian\/} metric tensor, then
\begin{equation}
N\cap\bar{N}=0\quad\text{so that}\quad \mathbb C\otimes TM=N\oplus 
\bar{N} \label{amt:pR}
\end{equation}
and one can define an orthogonal 
{\em almost complex\/} structure \( J \) on \( M \) by 
putting 
\begin{equation}
J(n+\bar{n})=\ri (n-\bar{n}) \quad\text{for every}\quad n\in N.
\label{amt:ac}
\end{equation} 
\smallskip

If \( g \) is {\em neutral\/} (i.e. of signature \( (2,2) \)), then 
there are two possibilities: either (\ref{amt:pR}) holds and there is 
an orthogonal, almost complex structure on \( M \) or
 \begin{equation}
 	N=\bar{N}\quad\text{so that}\quad N=\mathbb C\otimes K,
 	\label{amt:neutral}
 \end{equation} 
 where \( K=K^{\perp} \) is now a real, totally null, {\em plane\/} subbundle  of
 \( TM \).
 \smallskip
 
 In every one of the above cases, the complex, totally null, plane bundle \( N \) can be 
 characterized, at least locally, by (the direction of)
  a complex, decomposable 2-form \( F \) such 
 that
 \begin{equation}
 	n\in N\quad\text{iff}\quad n\in\mathbb C\otimes 
 	TM\quad\text{and}\quad n\lrcorner\vts F=0.
 	\label{amt:defF}
 \end{equation}
 If \( n_{1} \) and \( n_{2}\in\varGamma(N) \) are linearly 
 independent, then one can take \( F=g(n_{1})\wedge g(n_{2}) \). If \( F \)
 corresponds, in the sense of (\ref{amt:defF}), to \( N \), then \( 
 \star F \) corresponds to \( N^{\perp}\); since \( N^{\perp}=N \), the forms
 \( F \) and \( \star F \) are parallel. Using the notation of 
 (\ref{amt:iota}) one has \( \star F=\pm\iota F \).
 In signature \( (3,1) \) one 
 has 
 \( F\wedge \bar{F}=0 \) since \( F \) and \( \bar{F} \) have \( g(k) 
 \) as a common factor; in the other 
 two signatures, 
 if (\ref{amt:pR}) holds and \( F\neq 0 \), then \( F\wedge 
 \bar{F}\neq 0 \). 
 
 There is also a convenient, spinorial description of the bundles \( N 
 \). Assume, for simplicity, that there is a spin structure 
 \( Q \) on \( M \); spinor and tensor fields can be then represented 
 by equivariant maps from \( Q \) to suitable representation spaces;
 for example, a spinor field is given by a map \( \varphi:Q\to S \) 
 such that \( \varphi(qa)=\gamma(a^{-1})\varphi(q)\) for \( q\in Q \) 
 and \( a \) in the spin structure group of the bundle \( Q\to M \). 
 Given a totally null plane bundle \( N \) on a Riemannian 4-manifold, 
  there is a (locally defined) Weyl spinor field \( 
 \varphi \) on \( M \) such that
 \begin{equation}
 	N=\{n\in\mathbb C\otimes TM:\gamma(n)\varphi=0\}.
 	\label{amt:Nf}
 \end{equation}
 The chiralities of \( \varphi \) and \( F \) coincide: if \( 
 \gamma_{5}\varphi=\iota\varphi \), then for the corresponding 
 2-form \( F \) one has \( \star F=\iota F \). 
The isomorphisms \( \kappa \) and \( B \) of Section \ref{amt:CHK}, 
together with the representation \( \gamma \), 
induce, in dimension 4, an isomorphism of \(S_{+}\otimes_{\rm sym}S_{+}  
\) onto the complex space \( \bwedge^{2}_{+}\mathbb C^{4} \) 
of 2-forms \( F \) which are 
self-dual in the sense that \(\star F=\iota F  \); there is a similar 
isomorphism for spinors and 2-forms of the opposite chirality; these 
isomorphisms establish a correspondence between the descriptions of
\( N \) by means of 2-forms and spinors \cite{PR}. In the Lorentzian 
case, the  product \( \varphi \otimes \varphi_{c}\) corresponds to \( 
k\in\varGamma(K) \); in the proper Riemannian and neutral cases, if
\( \langle B\varphi_{c},\varphi\rangle\neq 0 \), then 
\( \varphi \otimes \varphi_{c}/\langle B\varphi_{c},\varphi\rangle \) 
corresponds to \( J \); in the neutral case, if 
\( \langle B\varphi_{c},\varphi\rangle=0 \), then \( \varphi \) is 
(proportional to) a Weyl-Majorana spinor.

In the Lorentzian case, the real part of  the 2-form \( F \) can be 
interpreted as  representing a `null'  electromagnetic field 
\( ({\mathbf E}, {\mathbf B}) \), i.e. a field such that 
\( ({\mathbf E}+\ri{\mathbf B})^{2}=0 \). 
In the 1950s, Ivor Robinson considered solutions of Maxwell's 
equations
 \begin{equation}
 	\rd F=0
 	\label{amt:Max}
 \end{equation}
 for such a null field and has shown that the trajectories of the flow 
 generated by \( k\in\varGamma(K) \) constitute a congruence of null 
 geodesics without shear \cite{R}. He  conjectured also that, given any 
 such smooth congruence on a Lorentzian manifold,
  one can find a non-zero solution \( F \) of 
  (\ref{amt:Max}) such that \( \star F=\ri F \) and \( k\lrcorner\vts 
  F=0 \). In 1985, Jacek Tafel \cite{Ta} pointed out that this need 
  not be true,  because, to find such a solution, one has to solve  a 
  linear, partial  differential equation of the first order, \( 
  \varLambda f=a\),
   of the type considered by Hans Lewy \cite{HL} and shown by 
  that author not  to have solutions, even locally, 
   for some smooth, but non-analytic, 
  functions \( a \); see also \cite{N}. Soon afterwards, it became clear \cite{RT2} that the
  structure underlying   shearfree congruences of null geodesics on 
  Lorentzian manifolds is that of  Cauchy-Riemann manifolds, earlier introduced 
  into physics by Penrose, in his theory of twistors associated with 
  Minkowski space, and its generalization to curved manifolds   
  \cite{Pe}-\cite{PR}.

 \begin{proposition} \label{prop3}
 Let \( M \) be a Riemannian 4-manifold with  a metric tensor \( g \) 
 that is either proper Riemannian or Lorentzian or neutral.
 Let \( N\to M \) be a totally null, complex, plane subbundle of \( 
 \mathbb C\otimes TM \) and let \( F \) be a 2-form such that 
 {\rm (\ref{amt:defF})} holds. Then
 
 \noindent{\ \ \rm (i)} equation {\rm (\ref{amt:Max})} implies the 
 {\em complex integrability condition:}
 \begin{equation}
 	[\varGamma(N),\varGamma(N)]\subset \varGamma(N);
 	\label{amt:GG}
 \end{equation}
 
 \noindent{\ \rm (ii)} if \( N\cap\bar{N}=0 \), then {\rm 
 (\ref{amt:GG})} is equivalent to the integrability of the almost 
 complex structure \( J \) defined by {\rm (\ref{amt:ac})}; if \( g \) 
 is proper Riemannian (resp., neutral), then {\rm (\ref{amt:defh})} 
 defines a proper Hermitian (resp., Hermitian of signature (1,1)) 
 tensor field \( h \) on \( M \);
 
 \noindent{\rm (iii)} if \( g \) is Lorentzian, then {\rm 
 (\ref{amt:GG})} is equivalent to the statement that the trajectories 
 of the flow generated by every \( k\in\varGamma(K) \), \( K \) as in
 {\rm (\ref{amt:NK})}, constitute a congruence of  {\em null geodesics 
 without shear}; moreover, if the congruence is regular in the sense 
 that the quotient set \( M'=M/K \) is a 3-manifold and the map
 \(\pi:M\to M'\)
 is a submersion, then \( N \) 
 projects to a complex line bundle \( H\to M' \), \( H\subset\mathbb 
 C\otimes TM' \), defining a {\em CR-structure} on \( M' \); the form 
 \( F \) satisfying {\rm (\ref{amt:defF})} and {\rm (\ref{amt:Max})} descends
 to a complex 2-form \( F' \) on \( M' \) such that  
 \begin{equation}
 	\rd F'=0,\quad Z\lrcorner\vts F'=0\;\; \text{for every}\;\;
 	Z\in H\quad\text{and}\quad F=\pi^{*}F';
 	\label{amt:M'}
 \end{equation}
 
 \noindent{\rm (iv)} if \( g \) is neutral and {\rm 
 (\ref{amt:neutral})} holds, then {\rm (\ref{amt:GG})} reduces to the 
 {\em real integrability condition,}
 \[
 	[\varGamma(K),\varGamma(K)]\subset \varGamma(K);
\]
the leaves of the foliation defined by \( K \) are 2-dimensional, 
totally null and totally geodesic  submanifolds of \( M \). 
 \end{proposition}
 The proof of Prop. \ref{prop3} is straightforward; most of it can can be found
 in \cite{HM,Nu,PR,RT,T1,T2}. There are interesting results and problems 
 connected with the analogy between  a shearfree congruence of null 
 geodesics on a Lorentz manifold and the Hermitian geometry in the 
 proper Riemannian case; one of them consists in 
 the proof of the Goldberg-Sachs theorem in signatures \( (4,0) \) 
 and \( (2,2) \) \cite{PB}.
 
 It is worth noting that, in the 
 Lorentzian case, the complex structure in the fibres of \( K^{\perp}/K 
 \) is determined, in a natural manner, by giving only a space and time
  orientation of \( M \) and the bundle 
 of null lines \( K \); in the proper Riemannian case, the (almost) 
 complex structure has to be introduced explicitly,  by giving either  \( 
 J \) or \( N \). It is for this reason that the appearance of the Cauchy-Riemann 
 equations in \cite{RT1} had been somewhat unexpected.

Recall that the (abstract) Cauchy-Riemann structure on a 3-manifold \( M' \), 
given by the complex line bundle \( H\to M' \), \(H\subset \mathbb 
C\otimes TM'  \), \( H\cap\bar{H}=0 \), can be conveniently 
 locally described also as follows: 
let \( Z \) be a non-zero section of \( H \) and let \( 
\lambda\) be a non-zero, real 1-form on \(  M\) such that  \( 
Z\lrcorner\vts \lambda=0 \). One can find a complex 1-form \( \mu \) such that 
\( \lambda\wedge\mu\wedge\bar{\mu}\neq 0 \), \( Z\lrcorner\vts\mu= 0 \)
and \( Z\lrcorner\vts\bar{\mu}\neq 0 \). 
These forms are defined  up to transformations
\begin{equation}
	\lambda\mapsto a\lambda,\quad \mu\mapsto b\mu+c\lambda,
	\label{amt:change}
\end{equation}
where \( a \) is a real function and \( b,c \) are complex 
 functions on \( M' \) such that \( a,b\neq 0 \). The direction of the 
 2-form \( \lambda\wedge\mu \) is invariant with respect to the 
 changes (\ref{amt:change}) and  characterizes the CR structure.
 In the terminology of \cite{J}, such a form is a section of the
 {\em canonical bundle\/} of the CR 3-manifold, defined as
 \[
\{\omega\in \mathbb C\otimes 
\bwedge^{2}T^{*}M':Z\lrcorner\vts\omega=0\;\;\text{for every}\;\; Z\in H\}.
 \]
To alleviate the language, I shall use, from now on,  the expression
`a CR space' instead of `a three-dimensional manifold with a CR 
structure'.

 Consider the fibration \(\pi: M\to M' \). Let \( 
 P\) and \( \xi \) 
 be a  real function and a real one-form on \( M \), 
 respectively, such that \( P^{2}
 \pi^{*}(\mu\wedge\bar{\mu}\wedge\lambda)\wedge \xi \)  
 vanishes nowhere on \( M \).
 The symmetric tensor field on \( M \),
 \begin{equation}\label{amt:RT}
 g=P^{2}\pi^{*}(\mu\otimes_{\rm sym}\bar{\mu})+\pi^{*}\lambda\otimes_{\rm sym}\xi,
 \end{equation}
 is the most general Lorentz metric admitting the fibres of \( \pi \) 
 as null geodesics constituting a congruence without shear \cite{RT2}.
 One then has \( \pi^{*}\lambda\wedge g(k)=0\) for \( k\in\varGamma(K) 
 \) and \( K^{\perp}=\ker \pi^{*}\lambda \).
 
Let \( f:M'\to\mathbb C \)
 be a smooth function. If the {\em Cauchy-Riemann 
equation\/}
\begin{equation}
	Z\lrcorner\vts \rd f=0
	\label{amt:CR}
\end{equation}
has two independent  (local) solutions \( z \) and \( w \), then
 the 2-form \( \rd z\wedge\rd w \) is a non-zero section of the 
 canonical bundle; using the freedom 
implied by (\ref{amt:change}), one can choose \( \mu \) to coincide 
with \( \rd z \). 
The map
\begin{equation}
	(z,w):M'\to \mathbb C^{2}
	\label{amt:embed}
\end{equation}
is  a (local) embedding of \( M' \) in \( \mathbb C^{2} \) and the CR 
structure is then said to be (locally) {\em embeddable\/}.
Lewandowski, Nurowski and Tafel have shown that the CR space defined by 
a shearfree congruence of null geodesics on an Einstein-Lorentz 
manifold is so embeddable \cite{LNT}.
One can then introduce  local coordinates \( (u,x,y) \) on \( M' \)  
such that \( x+\ri y=z \) and represent the form \( \lambda \) and the 
vector field \( Z \) as
\[
\lambda=\rd u+\bar{L}\rd z+L\rd\bar{z},\quad 
Z=\frac{\partial}{\partial \bar{z}}-L\frac{\partial}{\partial u}.
\]
If \( L=0 \), then the CR structure is trivial in the sense that \( M' 
\) is foliated by a family of complex 1-manifolds of equation \( u= \)const.; the 
corresponding bundle \( K^{\perp}\) is integrable, \( \lambda\wedge\rd 
\lambda=0 \), and  (\ref{amt:CR}) reduces to the classical 
Cauchy-Riemann equation: this is the special case of a  {\em `hypersurface 
orthogonal'\/} congruence of shearfree null geodesics considered in 
\cite{RT1}.

According to part (iii) of Prop. \ref{prop3} 
the general problem of finding a solution of Maxwell's equations 
(\ref{amt:Max}) adapted to a shearfree congruence of null geodesics
defined by \( K \), i.e. such that \( \star F=\ri F \) and
\( k\lrcorner\vts F=0 \),  reduces to the 
following: given a CR space \( M' \), find a  closed  section   \( F' \) 
of its canonical bundle. 
If \( M' \) is embeddable, then such sections exist and are of the 
form
\[
F'=f(z,w)\rd z\wedge \rd w  
\]
where \( z \) and \( w \) are as in (\ref{amt:embed}) and \( f \) is 
an analytic function of its arguments.

It is now known that there are CR  spaces  
that are non-embeddable, but have one solution of (\ref{amt:CR}) \cite{Ro};
by the results of \cite{Ta}, extended to higher dimensions in 
\cite{J}, such CR spaces do not admit closed, non-zero sections of 
their canonical bundle.\footnote{The significance,  in this context, of the examples 
found by Rosay and of the results of Jacobowitz
 has been explained at the Workshop in Vienna by C. 
Denson Hill.}
Therefore, Lorentzian manifolds constructed  on the basis of these CR 
spaces as in (\ref{amt:RT}) do not admit any non-zero solutions \( F \) of 
Maxwell equations  such that \( k\lrcorner\vts F=0 \), \( g(k)\wedge 
F=0 \), where \( g(k) =\lambda\). 
There are examples of non-embeddable 7-dimensional CR manifolds that
have non-zero, closed, sections of their canonical bundle, but it is 
not clear whether there are such examples in dimensions 3 and 5.
In connection with this, I formulate the following 
\begin{conj} A CR 3-space admits locally a closed, non-zero section of its 
canonical bundle if, and only if, it is locally embeddable.
\end{conj}  
The conjecture can be   formulated as a problem
of elementary vector calculus: given a complex 
vector field \( \mathbf F \)  on \( \mathbb R^{3} \) such that 
\( \mathbf F\times\overline{\mathbf F}\neq 0 \) and \( {\rm 
div}\ts\mathbf F=0 \), show that there exist two 
complex functions \( z \) and \( w \) such that \( \mathbf 
F={\rm grad}\ts z\times {\rm grad}\ts w \).

Since it is known that real analytic CR spaces are locally embeddable, the 
proof of the conjecture---if it is true---should concern the smooth 
case.  

 \section*{Acknowledgments}

Work on this article was supported in part by the Polish Committee on Scientific 
Research (KBN) under grant no. 2 P03B 017 12 and by 
 the Foundation for Polish-German
cooperation with funds  provided by the Federal Republic of Germany. 
The paper has been completed in June 1997, during the Workshop on
{\sl  Spaces of geodesics and complex structures in general 
relativity\/} at the Erwin Schr\"odinger International Institute for 
Mathematical Physics in Vienna. 
I have benefited there from 
discussions with C. Denson Hill,  Pawe{\l} Nurowski, Roger Penrose and 
Helmuth Urbantke.


\begin{thebibliography}{99}
\bibitem{BT} Budinich, P. and Trautman, A., {\em The spinorial 
chessboard\/}, Springer-Verlag, Berlin 1988.
\bibitem{Za} Ebbinghaus, H.-D., Hermes, H., Hirzebruch, F., Koecher, 
M., Mainzer, K., Prestel, A. and Remmert, R., {\em Zahlen\/}, 
Springer-Verlag, Berlin 1983.
\bibitem{EK} Ehlers, J. and Kundt, W., Exact solutions of the 
gravitational field equations, in: {\em Gravitation\/}, edited by L. 
Witten, Wiley, New York 1962.
\bibitem{HM} 
 Hughston, L. P.  and  Mason, L. J.,   A generalised
Kerr--Robinson theorem, {\em  Class. Quantum Grav.} {\bf 5} (1988) 275--285.
\bibitem{J} Jacobowitz, H., The canonical bundle and realizable CR 
hypersurfaces, {\em Pacific J. Math.\/} {\bf 127} (1987) 91--101.
\bibitem{LNT}  Lewandowski, J.,  Nurowski, P., and Tafel, J.,  Einstein's
equations and realizability of CR manifolds, {\em Class. Quantum Grav.}
{\bf 7} (1990) L241 - L246.
\bibitem{HL} Lewy, H., An example of a smooth partial differential 
equation without solution, {\em Ann. of Math.\/} {\bf 66} (1957) 
155--158.
\bibitem{Lo} Lounesto, P., Counter-examples in Clifford algebras,
{\em Adv. Appl. Clifford Algebras\/} {\bf 6} (1996) 69--104.
\bibitem{N} Nirenberg, L., On a question of Hans Lewy, {\em Russian 
Math. Surveys\/} {\bf 29} (1974) 251--262.
\bibitem{Nu} Nurowski, P., Optical geometries and related structures,
{\em J. Geom. Phys.\/} {\bf 18}  (1996) 335--348.
\bibitem{Pe} Penrose, R., The complex geometry of the natural world, in:
{\em Proc. Int. Congress Math.}, pp. 189--194, Helsinki 1978. 
\bibitem{Pn} Penrose, R., Physical space-time and nonrealizable 
CR-structures, {\em Proc. Symp. Pure Math.\/} {\bf 39}, Part I (1983) 
401--422.  
\bibitem{PR} Penrose, R. and Rindler, W., {\em Spinors and 
space-time\/},
vols 1 and 2, Cambridge University Press, Cambridge  1984 and 1986.
\bibitem{PB} Przanowski, M. and Broda, B., Locally K\"ahler 
gravitational instantons, {\em Acta Phys. Polon.\/} {\bf B14} (1983) 
637--661.  
\bibitem{R} Robinson, I.,  Null electromagnetic fields, {\em J.
Math. Phys.\/} {\bf 2} (1961) 290--291.
\bibitem{RT1} Robinson, I. and Trautman, A., 
Spherical gravitational waves, 
{\em Phys. Rev. Lett.\/} {\bf 4} (1960) 431--432.
\bibitem{RT2}  Robinson, I. and  Trautman, A., Cauchy--Riemann
structures in optical geometry  in: {\em
Proc. of the Fourth Marcel Grossmann Meeting on General Relativity}, 
pp. 317--324, 
edited by R. Ruffini, Elsevier, 1986. 
\bibitem{RT} Robinson, I. and Trautman, A., Optical geometry, in:
{\em New Theories in Physics\/}, Procs. of the XI Warsaw Symposium on 
Elementary Particle Physics, edited by Z. Ajduk {\em et al.\/}, World 
Scientific, Singapore 1989.
\bibitem{Ro} Rosay, J.-P., New examples of non-locally embeddable CR 
structures, {\em Ann. 
Inst. Fourier Grenoble\/} {\bf 39} (1989) 811--823.
\bibitem{So}  Sommerfeld, A., Mathematische Theorie der Diffraction, 
{\em Math. Annalen\/} {\bf 47} (1896) 317--374.
\bibitem{Ta} Tafel, J., On the Robinson theorem and shearfree 
geodesic null congruences, {\em Lett. Math. Phys.\/} {\bf 10} (1985) 
33--39.
\bibitem{Th} Thirring, W., {\em Principles of quantum 
electrodynamics\/}, Academic Press, New York 1958.
\bibitem{T}
Trautman, A., Optical structures in relativistic theories, in: {\em  \'Elie Cartan et les
math\'emathiques  d'aujourd'hui\/}, {\em
Ast\'erisque}, num\'ero hors s\'erie, (1985) 401--420.
\bibitem{T1} Trautman, A., Geometric aspects of spinors, in: {\em 
Clifford algebras and their applications in mathematical physics\/}, 
edited by  R. Delanghe, F. Brackx and H. Serras, Kluwer Academic 
Publishers, Dordrecht 1993.
\bibitem{T2} Trautman, A., Clifford and the `square root' ideas, 
{\em Contemporary Mathematics\/}, {\bf 203} (1997) 3--24. 
\end{thebibliography}
\end{document}